# Mortgage-Rate-Adjusted Home Prices

Honggao Cao[1]
Wells Fargo & Company
June 30, 2022

In this paper, we investigate the impact of mortgage rates on home prices, and how the impact may be used to help property-purchase discussions at individual buyer level and to adjust home price indices across time. A mortgage-rate-adjusted "effective price" is derived to measure near-term property price in the presence of (*expected*) mortgage rate changes. A price-mortgage rate neutrality line is then constructed based on the "effective price" to help differentiate various market scenarios in the near term, which can be used by prospective buyers in their "to-buy or not-to-buy" deliberations. At the market level, effective home prices allow for neutralization of the mortgage rates on the movement of the housing market. An application of the neutralization strategy to the Case-Shiller Home Price Index (HPI) indicates that the U.S. housing market has been considerably affected by the dynamics of mortgage rates in a long run. But mortgage rates do not appear a primary driver of the extraordinary home price increase during the COVID-19 pandemic.

## A. Introduction and Conclusions

Mortgage rates and home prices are intriguingly related. This is the case because home purchases are predominantly supported by mortgage financing. According to National Association of Realtors (NAR, [1]), for example, 87% of home buyers in 2021 financed their home purchases through mortgages; of those who used mortgages, first-time buyers typically financed 93% of their purchases whereas repeat buyers financed 83%. The relationship between mortgage rates and home prices is complex, however. First, home prices are affected by many factors ( [2], [3], [4], [5], [6], [7], [8]); mortgage rates are just one of them. Second, mortgage rates are "endogenous" as they are often an important channel through which changes in the government's monetary policy affect consumer balance sheets and spending, including spending and thus prices in the housing sector ( [9] and [10]). This may make the relationship between mortgage rates and home prices circular ( [11]).

The complexity in the relationship between mortgage rates and home prices can create difficulties for prospective home buyers as well as other stakeholders, especially when mortgage rates are actively used as a policy tool. Should they buy now given the prevailing mortgage rates and home prices, or wait to take advantage of an expected price drop as a result of an expected mortgage rate increase? If they buy into a growing housing market, how much caution should they use to account for a potential market reverse from a potential mortgage rate increase? More broadly, given ever changing mortgage rates, how should the prevailing home prices be evaluated as part of home purchase consideration?

In this paper, we present an analytical framework that may help answer the above questions. Specifically, we argue that nominal home prices *alone* are not a clear or complete measure of housing market conditions. Just like an old saying that all housing markets are local (and thus different *spatially*), home prices are "local"

---

[1] Opinions expressed in this paper do not necessarily reflect those of Wells Fargo and Company. I thank Stewart Brown and Yi Su for comments on a previous version of the paper. All errors are my own. Correspondence should be addressed to honggao.cao@wellsfargo.com.

*temporally* as the same prices may have different financial implications for home buyers when mortgage rates change across time. Based on this temporal price locality, we construct an alternative measure of "effective" home prices by examining a stylized, representative home purchasing process, explicitly adjusting for the impact of mortgage rates on the purchases (Sections B and C). Using this alternative measure, we then derive a price-mortgage rate neutrality line to differentiate various market scenarios in the near term, which can be used to help prospective buyers in their "to-buy or not-to-buy" deliberations (Sections D and E).

At the market level, we show that the effective home prices allow for neutralization of the mortgage rates on the movement of the general housing market. An application of the neutralization strategy to the Case-Shiller Home Price Index (HPI) indicates that the U.S. housing market has been considerably affected by the dynamics of mortgage rates in a long run (Section F). But mortgage rates do not appear a primary driver of the extraordinary home price increase during the COVID-19 pandemic (Section G).

It should be clearly noted that our analytical framework does not predict future home prices, or capture the full complexity of the housing market. But by focusing on some key features of the home purchasing process, and by considering the impact of mortgage rates in the process, the framework offers a useful filtering tool for processing the price information in the housing market. Home buyers and other relevant stakeholders are advised that home prices and mortgage rates should always be viewed in tandem in the housing market-related decisions and discussions.

## B. Contemporaneous Purchases

Consider two compatible properties purchased at the same time. Property A is purchased at price $P_A$, and Property B at $P_B$.

Each purchase is financed through a fixed-term mortgage, with a down payment that is proportional to the purchase price. The mortgage term ($T\ months$) and down payment rate ($\alpha$) are assumed identical between the two purchases, but the mortgage rates, $r_A$ and $r_B$, are not. [2]

The scheduled monthly payments for the two purchases, $X_A$ and $X_B$, are as follows:

$$X_A = (1-\alpha)\frac{r_A(1+r_A)^T}{(1+r_A)^T-1}P_A \qquad (1)$$

$$X_B = (1-\alpha)\frac{r_B(1+r_B)^T}{(1+r_B)^T-1}P_B \qquad (2)$$

The down payments required for the purchases, $D_A$ and $D_B$, are:

$$D_A = \alpha P_A \qquad (3)$$

$$D_B = \alpha P_B \qquad (4)$$

---

[2] In this setup, we are focused on home purchases through fixed-term mortgages alone. In some of the ensuing illustrative analysis, we further restrict our attention to 30-year fixed rate mortgages. While this focus and restriction does not capture the full complexity of the housing markets (e.g., cash purchases, variable rate mortgages, and 15-year fixed rates), it covers a significant portion (e.g., 80+%) of the housing market activities ( [12], [13] and [14]).

The purchase *or nominal* prices for the two properties may be viewed as the sum of the down payments and the discounted payment flows, as seen in the following scheme:

$$P_A = \frac{(1+r_A)^T - 1}{r_A(1+r_A)^T} X_A + D_A \equiv \beta_A X_A + D_A = \frac{(1+r_A)^T - 1}{r_A(1+r_A)^T} * (1-\alpha) \frac{r_A(1+r_A)^T}{(1+r_A)^T - 1} P_A + \alpha P_A$$

$$= (1-\alpha)\beta_A \frac{1}{\beta_A} * P_A + \alpha P_A = P_A \tag{5}$$

$$P_B = \frac{(1+B)^T - 1}{r_B(1+r_B)^T} X_B + D_B \equiv \beta_B X_B + D_B = \frac{(1+r_B)^T - 1}{r_B(1+r_B)^T} * (1-\alpha) \frac{r_B(1+r_B)^T}{(1+r_B)^T - 1} P_B + \alpha P_B$$

$$= (1-\alpha)\beta_B \frac{1}{\beta_B} * P_B + \alpha P_B = P_B \tag{6}$$

Here, the βs are discounting factors corresponding to a given mortgage rate and mortgage term, a *non-negative* metric that is essential to the discussion in the rest of this paper. For the ease of exposition, we formally define the discounting factor as follows:

$$\beta(r;T) = \frac{(1+r)^T - 1}{r(1+r)^T} \equiv \beta(r) \tag{7}$$

The mortgage term ($T$) is removed from the definition of β as we assume, for practical reasons, that all mortgages share the same $T$'s.

**C. Sequential Purchases**

Let us now change our setup and assume that Properties A and B are purchased sequentially, with A at time $t$, and B at $t+1$. In this new setup, the difference between the mortgage rates, $r_A$ and $r_B$ – if any – can be seen as changes in mortgage rate across time.

Relative to time $t$ (or given mortgage rate $r_A$), the purchase price for Property B becomes

$$P_B^A = \beta_A X_B + D_B = \frac{(1+r_A)^T - 1}{r_A(1+r_A)^T} * (1-\alpha) \frac{r_B(1+r_B)^T}{(1+r_B)^T - 1} P_B + \alpha P_B$$

$$= \{(1-\alpha) \frac{(1+r_A)^T - 1}{r_A(1+r_A)^T} * \frac{r_B(1+r_B)^T}{(1+r_B)^T - 1} + \alpha\} P_B$$

$$= \{(1-\alpha) \frac{\beta_A}{\beta_B} + \alpha\} P_B$$

$$\equiv \gamma(r_A, r_B; T, \alpha) P_B \tag{8}[3]$$

This adjusted price for property B can be interpreted as the *effective* price at time $t$ for a potential purchase of a compatible property at the next period – if the nominal purchase price is $P_B$ and mortgage rate is expected to change from $r_A$ to $r_B$. As detailed later, this is a price that one can use to gauge the relative preferability of two alternative property purchase decisions: to purchase now at price $P_A$ and mortgage rate $r_A$, or to purchase in next period at price $P_B$ and mortgage rate $r_B$.

---

[3] In this formulation, the potential earnings from delayed down payment are assumed negligible. This assumption is reasonable for all practical purposes in this paper, as the effective price is always evaluated relative to one period prior (e.g., $t+1$ versus $t$). Alternative formulations should be considered to explicitly account for the effect of delayed down payment when the time gap between the purchases is significant.

The effective price contains a multiplicative adjuster, $\gamma$. Given the mortgage term ($T$) and the down payment rate ($\alpha$), this mortgage-rate-based price adjuster captures all the effect of mortgage rate change on the purchase of Property B.

The behavior of $\gamma$ warrants some attention. First, $\gamma$ is always positive, as both $\beta_A$ and $\beta_B$ are positive. This guarantees that the effective price of any future property can never go below 0, regardless of mortgage rate. In addition, $\gamma$ will be greater than 1 when the next period mortgage rate $r_B$ is greater than the baseline rate $r_A$, and vice versa. This suggests that other things being equal, the effective property price moves in the same direction as the relative mortgage rate: A future property purchase will *effectively* be more expensive than suggested by the nominal price when mortgage rate increases, and less so when mortgage rate decreases.

### D. Effective Price Changes

The difference between the nominal and effective prices in the presence of mortgage rate changes as seen in equation (8) can be used to measure the effective price change across time. Let the purchase or nominal price growth rate be $g = \frac{P_B - P_A}{P_A}$. The effective, mortgage-rate-adjusted growth rate will then be

$$g^* = \frac{P_B^A - P_A}{P_A} = \gamma(1+g) - 1 \qquad (9)$$

To see how the effective and nominal price growth rates are related to each other under different mortgage rate regimes, consider the following scenario: the baseline mortgage rate $r_A$ = 4.5%, m*ortgage term in month T* = 360, and the down payment ratio $\alpha$ = 20%. The behavior of the effective price growth rate for selected nominal rates (g) under different mortgage rate assumptions (r) is illustrated in the following table. By design, when the mortgage rate is at the baseline rate (4.5%, highlighted in yellow), the effective rates and the nominal rates are identical. When mortgage rate decreases and becomes lower than the baseline, the effective rates will be lower than the nominal rates. Conversely, when mortgage rate increases and becomes higher than the baseline, the effective rates will be higher than the nominal.

**Table 1. Mortgage-Rate Adjusted Effective Home Price Growth Rates**
(Baseline Mortgage Rate = 4.5%, T = 360 months, $\alpha$ = 20%; Positive Effective Rates are in red)

| Mortgage rate (r) | Nominal Growth Rate (g) | | | | | | | | | | | | |
|---|---|---|---|---|---|---|---|---|---|---|---|---|---|
| | -10% | -8% | -6% | -4% | -2% | -1% | 0% | 1% | 2% | 4% | 6% | 8% | 10% |
| 3.500% | -0.182 | -0.164 | -0.146 | -0.127 | -0.109 | -0.100 | -0.091 | -0.082 | -0.073 | -0.055 | -0.036 | -0.018 | 0.000 |
| 3.625% | -0.172 | -0.154 | -0.135 | -0.117 | -0.098 | -0.089 | -0.080 | -0.071 | -0.062 | -0.043 | -0.025 | -0.006 | 0.012 |
| 3.750% | -0.162 | -0.143 | -0.125 | -0.106 | -0.087 | -0.078 | -0.069 | -0.059 | -0.050 | -0.032 | -0.013 | 0.006 | 0.024 |
| 3.875% | -0.152 | -0.133 | -0.114 | -0.095 | -0.076 | -0.067 | -0.058 | -0.048 | -0.039 | -0.020 | -0.001 | 0.018 | 0.037 |
| 4.000% | -0.142 | -0.123 | -0.103 | -0.084 | -0.065 | -0.056 | -0.046 | -0.037 | -0.027 | -0.008 | 0.011 | 0.030 | 0.049 |
| 4.125% | -0.131 | -0.112 | -0.093 | -0.073 | -0.054 | -0.044 | -0.035 | -0.025 | -0.015 | 0.004 | 0.023 | 0.042 | 0.062 |
| 4.250% | -0.121 | -0.101 | -0.082 | -0.062 | -0.043 | -0.033 | -0.023 | -0.014 | -0.004 | 0.016 | 0.035 | 0.055 | 0.074 |
| 4.375% | -0.111 | -0.091 | -0.071 | -0.051 | -0.031 | -0.022 | -0.012 | -0.002 | 0.008 | 0.028 | 0.048 | 0.067 | 0.087 |

| 4.500% | -0.100 | -0.080 | -0.060 | -0.040 | -0.020 | -0.010 | 0.000 | 0.010 | 0.020 | 0.040 | 0.060 | 0.080 | 0.100 |
|---|---|---|---|---|---|---|---|---|---|---|---|---|---|
| 4.625% | -0.089 | -0.069 | -0.049 | -0.029 | -0.008 | 0.002 | 0.012 | 0.022 | 0.032 | 0.052 | 0.072 | 0.093 | 0.113 |
| 4.750% | -0.079 | -0.058 | -0.038 | -0.017 | 0.003 | 0.013 | 0.024 | 0.034 | 0.044 | 0.065 | 0.085 | 0.106 | 0.126 |
| 4.875% | -0.068 | -0.047 | -0.027 | -0.006 | 0.015 | 0.025 | 0.036 | 0.046 | 0.056 | 0.077 | 0.098 | 0.118 | 0.139 |
| 5.000% | -0.057 | -0.036 | -0.015 | 0.006 | 0.027 | 0.037 | 0.048 | 0.058 | 0.069 | 0.089 | 0.110 | 0.131 | 0.152 |
| 5.125% | -0.046 | -0.025 | -0.004 | 0.017 | 0.038 | 0.049 | 0.060 | 0.070 | 0.081 | 0.102 | 0.123 | 0.144 | 0.166 |
| 5.250% | -0.035 | -0.014 | 0.008 | 0.029 | 0.050 | 0.061 | 0.072 | 0.083 | 0.093 | 0.115 | 0.136 | 0.158 | 0.179 |
| 5.375% | -0.024 | -0.003 | 0.019 | 0.041 | 0.062 | 0.073 | 0.084 | 0.095 | 0.106 | 0.127 | 0.149 | 0.171 | 0.193 |
| 5.500% | -0.013 | 0.009 | 0.031 | 0.053 | 0.075 | 0.086 | 0.096 | 0.107 | 0.118 | 0.140 | 0.162 | 0.184 | 0.206 |
| 5.625% | -0.002 | 0.020 | 0.042 | 0.065 | 0.087 | 0.098 | 0.109 | 0.120 | 0.131 | 0.153 | 0.175 | 0.198 | 0.220 |
| 5.750% | 0.009 | 0.032 | 0.054 | 0.077 | 0.099 | 0.110 | 0.121 | 0.133 | 0.144 | 0.166 | 0.189 | 0.211 | 0.234 |
| 5.875% | 0.021 | 0.043 | 0.066 | 0.089 | 0.111 | 0.123 | 0.134 | 0.145 | 0.157 | 0.179 | 0.202 | 0.225 | 0.247 |
| 6.000% | 0.032 | 0.055 | 0.078 | 0.101 | 0.124 | 0.135 | 0.147 | 0.158 | 0.170 | 0.192 | 0.215 | 0.238 | 0.261 |
| 6.125% | 0.043 | 0.067 | 0.090 | 0.113 | 0.136 | 0.148 | 0.159 | 0.171 | 0.183 | 0.206 | 0.229 | 0.252 | 0.275 |
| 6.250% | 0.055 | 0.078 | 0.102 | 0.125 | 0.149 | 0.160 | 0.172 | 0.184 | 0.196 | 0.219 | 0.242 | 0.266 | 0.289 |
| 6.375% | 0.067 | 0.090 | 0.114 | 0.138 | 0.161 | 0.173 | 0.185 | 0.197 | 0.209 | 0.232 | 0.256 | 0.280 | 0.304 |
| 6.500% | 0.078 | 0.102 | 0.126 | 0.150 | 0.174 | 0.186 | 0.198 | 0.210 | 0.222 | 0.246 | 0.270 | 0.294 | 0.318 |
| 6.625% | 0.090 | 0.114 | 0.138 | 0.163 | 0.187 | 0.199 | 0.211 | 0.223 | 0.235 | 0.259 | 0.284 | 0.308 | 0.332 |
| 6.750% | 0.102 | 0.126 | 0.151 | 0.175 | 0.200 | 0.212 | 0.224 | 0.236 | 0.249 | 0.273 | 0.298 | 0.322 | 0.346 |
| 6.875% | 0.113 | 0.138 | 0.163 | 0.188 | 0.212 | 0.225 | 0.237 | 0.250 | 0.262 | 0.287 | 0.311 | 0.336 | 0.361 |
| 7.000% | 0.125 | 0.150 | 0.175 | 0.200 | 0.225 | 0.238 | 0.250 | 0.263 | 0.275 | 0.300 | 0.325 | 0.350 | 0.375 |

### E. Neutrality in Nominal Price and Mortgage Rate

The sign of the effective price growth rates contains important information. Relative to the baseline mortgage rate and current property price (or the "$g = 0$" column in the table), a negative effective growth rate indicates that a purchase at the alternative mortgage or price level is cheaper than suggested by the nominal rate, whereas a positive effective growth rate indicates that the alternative is more expensive.

The clear separation between positive (marked in red) and negative effective growth rates in the (r-g) table above indicates that in a world of changing mortgage rates and property prices, some alternative purchases can be clearly better off than others. If the current mortgage rate is treated as the baseline, the separation between positive and negative effective growth rates can also help evaluate if one should favor property purchase now or delay purchase to the future.

To demonstrate this, consider a property-price and mortgage-rate neutrality line in a r-g plane.[4] Given the baseline mortgage rate ($r_A$), and relative to the current price level (*i.e.*, $g = 0$), all purchases on this line are equivalent or "neutral" in a sense that the effective growth rate = 0. Purchases off this line with negative effective rates will be better (i.e., cheaper) purchases, and those with positive effective rates will be worse (or more expensive) purchases.

---

[4] As we will see in the ensuing discussion, technically, this should be termed "price change and mortgage rate change neutrality line". We use the term "price-mortgage rate neutrality" for simplicity.

Mechanically, let $g^*$ in equation (9) equal to 0,

$$g^* = \gamma(1+g) - 1 = 0 \rightarrow$$

$$g = \frac{1-\gamma}{\gamma} = \frac{(1-\alpha)(\beta_B - \beta_A)}{\alpha \beta_B + (1-\alpha)\beta_A} \tag{10}$$

Equation (10) describes the nominal price growth as a function of alternative mortgage rate on the neutrality line. Given a down payment rate, $\alpha$, it is completely determined by two mortgage-rate discounting factors, $\beta_A$ and $\beta_B$. The behavior of this neutrality line for four selected baseline mortgage rates ($r_A$) is plotted in Figure 1 below.

Figure 1. Price-Mortgage Rate Neutrality Lines for Selected Baseline Mortgage Rates

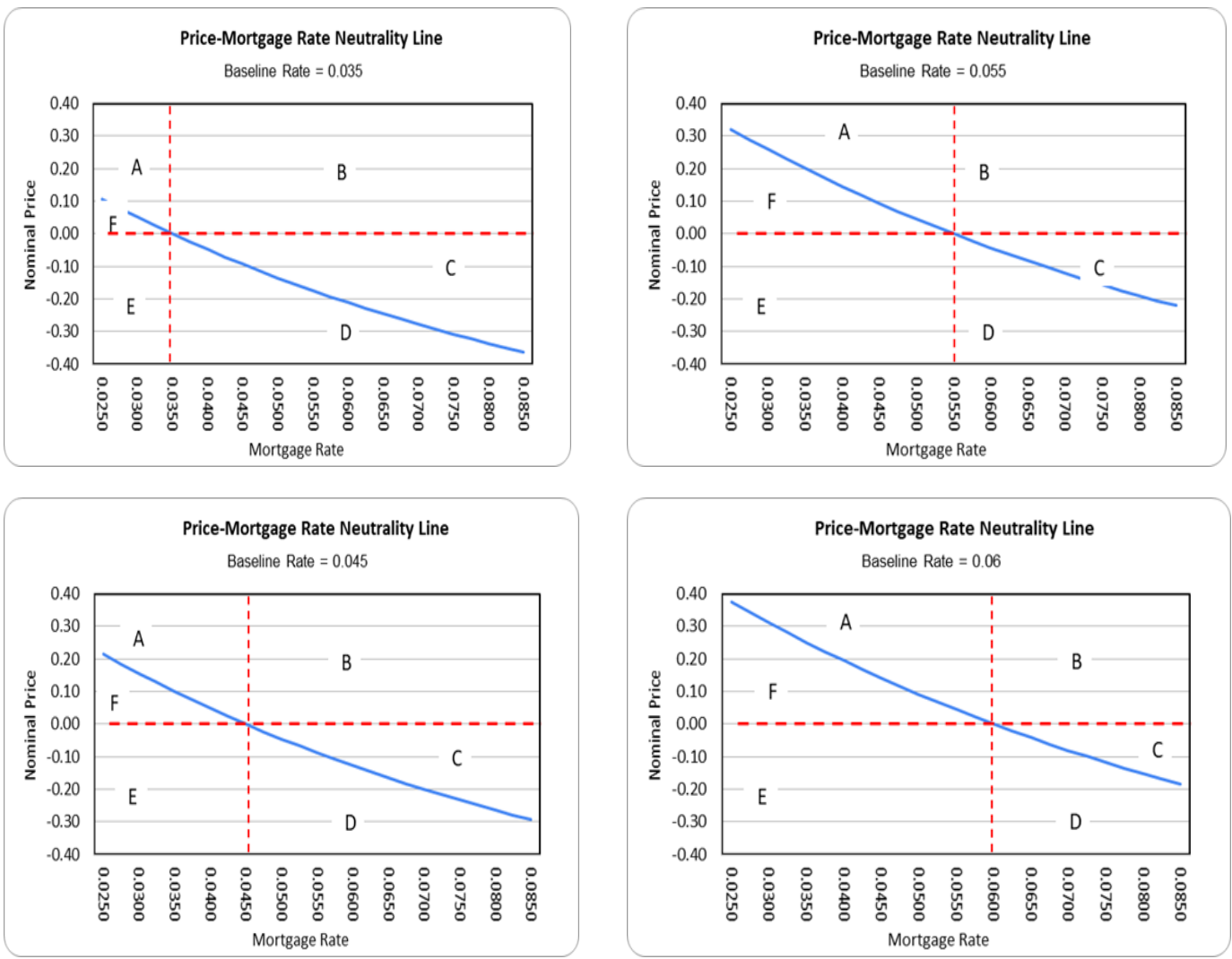

In each of the plots, the r-g plane is split into six regions by the baseline mortgage rate (vertical red line), current property price level (horizontal red) and the neutrality line. Regions D, E and F correspond to negative effective growth rates; any future purchase that falls into these regions will be better than the baseline purchase at the current mortgage rate and at the current price level. Conversely, any future purchase in regions A, B and C is worse than the baseline purchase. Some key characteristics of each region are summarized in Table 2 below.

Table 2. Alternative Property Purchase Decisions off the Price-Mortgage Rate Neutrality Line

| Region | Changes Relative to the Baseline | | Effective Price | Key Characteristics |
|---|---|---|---|---|
| | Nominal Price | Mortgage Rate | | |
| A | + | - | + | Should buy now; future mortgage rate reduction is not enough to compensate for price increase |
| B | + | + | + | Should buy now; double whamming from future increase in both mortgage rate and price |
| C | - | + | + | Should buy now; future price reduction is not enough to compensate for mortgage rate increase |
| D | - | + | - | Should wait; future mortgage rate increase is well compensated by price reduction |
| E | - | - | - | Should wait; future purchases benefit from reduction in both price and mortgage rate |
| F | + | - | - | Should wait; future price increase is well compensated by mortgage rate reduction |

## F.  Mortgage-Rate Neutral Home Price Index

The effective price and effective growth rate as determined in equations (8) and (9) can also be used to help derive mortgage-rate "neutral" home price index (HPI). Consider a generic HPI series, H = $\{h_1, h_2, ..., h_N\}$. If the corresponding mortgage rates over the period are captured as R = $\{r_1, r_2, ..., r_N\}$, then a mortgage-rate-adjusted, effective home price index may be formulated via the mechanism below:

i) From H, form a series of home price growth rate, G = $\{g_1 = 0, g_2, ..., g_N\}$, where $g_n = (h_n - h_{n-1})/h_{n-1}$.
ii) From R, form a series of mortgage-rate-based price adjuster, Γ = $\{\gamma_1 = 1, \gamma_2, ..., \gamma_N\}$, using the definition of $\gamma$ in equation (8).
iii) From G and Γ, form a series of effective home price growth rate, G* = $\{g_1^* = 0, g_2^*, ..., g_N^*\}$, using equation (9).
iv) From H and G*, form to a series of effective home price adjustment, K = $\{k_1 = 0, k_2, ..., k_N\}$, where $k_n = h_{n-1} g_n^*$.
v) From H and K, form a series of adjusted home price index, H* = $\{h_1^* = h_1, h_2^*, ..., h_N^*\}$, where $h_n^* = h_1 + \sum_{i=1}^{n} k_i$. Essentially, this final step collects price adjustment cumulatively across time to arrive at the mortgage-rate-adjusted HPI, and it can be shown

$$h_n^* = h_n + \sum_{i=1}^{n}(\Upsilon_i - 1)h_i \qquad (10)$$

Figure 2 below shows how the Case-Shiller HPI will look like if it is adjusted by the 30-year fixed mortgage rates and a 20% down payment assumption for four alternative timelines. The impact of the adjustment, including the effect of the down payment assumption, is summarized in Table 3.

From both the figure and the table, it is clear that the U.S. housing market has been considerably affected by the dynamics of mortgage rates. Regardless of the down payment rate assumptions, which appear to only affect the adjustment marginally, mortgage rate movements have contributed positively to the long-run U.S. housing market boom. Over a 35-year period from 1987 to 2021, for example, the mortgage-rate adjusted HPI would be about 24% lower; in the last three-years alone, mortgage rates have about 13% contribution to the home price growth.

**Figure 2. Mortgage-Rate-Adjusted Case-Shiller HPI by Alternative Timelines**

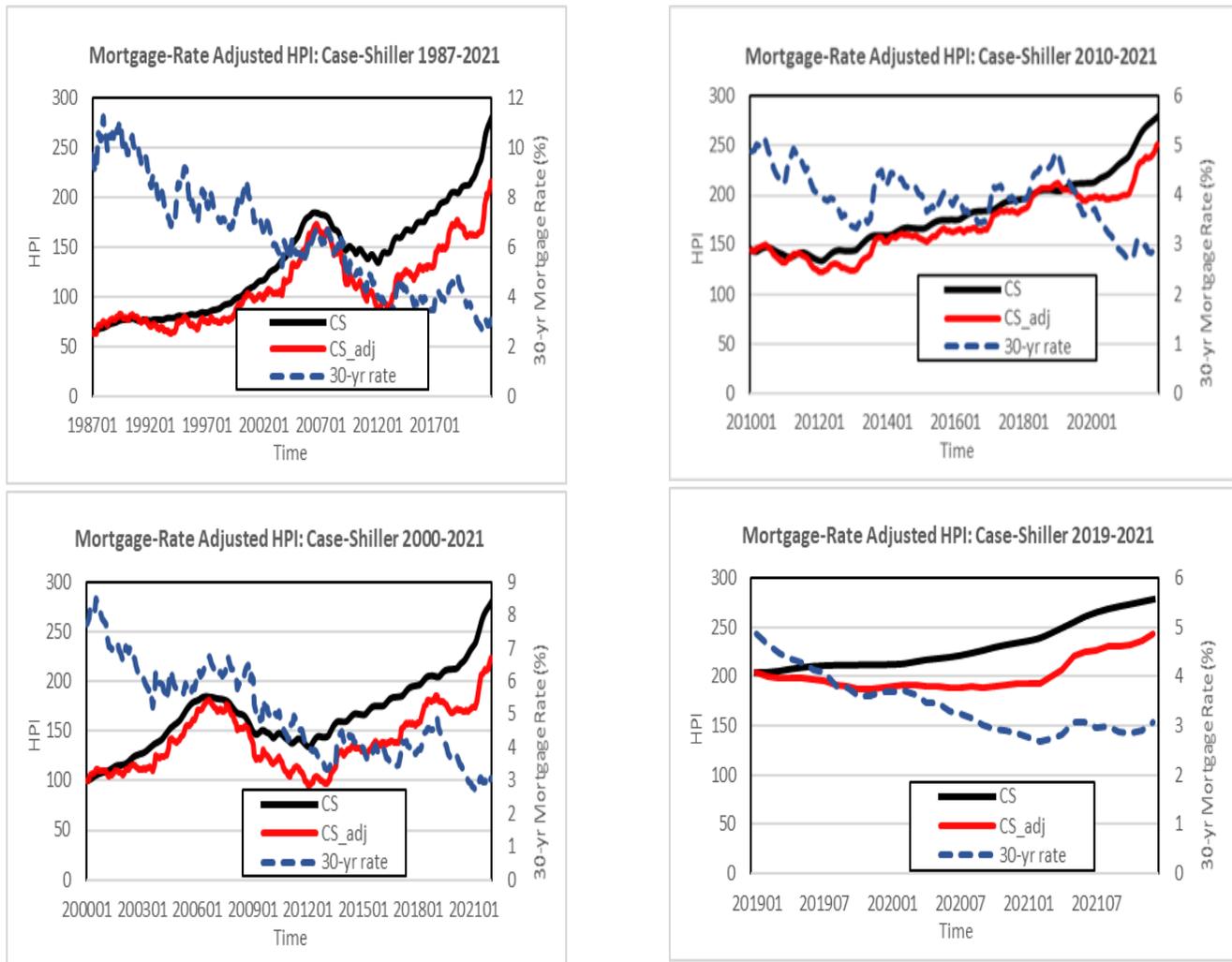

Note: Down payment rate is assumed as 20%. Mortgage rates are assumed to affect prices with a two-month lag. The HPI and mortgage rate data were retrieved from Federal Reserve Economic Data | FRED | St. Louis Fed (stlouisfed.org) on April 18, 2022

Table 3. The Impact of Mortgage Rates on Case-Shiller HPI

| Timeline | Down Payment Rate (α) | Nominal Growth (A) | Adjusted Growth (B) | Impact from Mortgage Rates (C) |
|---|---|---|---|---|
| 1987-2021 | 0.10 | 337.3% | 226.2% | 25.4% |
| | 0.15 | 337.3% | 232.4% | 24.0% |
| | 0.20 | 337.3% | 238.5% | 22.6% |
| 2000-2021 | 0.10 | 178.7% | 117.3% | 22.0% |
| | 0.15 | 178.7% | 120.7% | 20.8% |
| | 0.20 | 178.7% | 124.1% | 19.6% |
| 2010-2021 | 0.10 | 92.2% | 70.6% | 11.2% |
| | 0.15 | 92.2% | 71.8% | 10.6% |
| | 0.20 | 92.2% | 73.0% | 10.0% |
| 2019-2021 | 0.10 | 36.5% | 16.9% | 14.3% |
| | 0.15 | 36.5% | 18.0% | 13.5% |
| | 0.20 | 36.5% | 19.1% | 12.7% |

Note: C = (A-B)/(1+A).

## G. Home Prices during the Pandemic

The U.S. housing market has seen an extraordinary growth during the ongoing COVID-19 pandemic. Since the beginning of the pandemic, the Case-Shiller HPI has increased by about 30%, from 215.2 in Mach 2020 to 278.7 in December 2021. In the meantime, mortgage rates have stayed historically low for almost the entire pandemic – until the early 2022. The low mortgage rates, however, did not appear the primary reason for the huge run-up in home prices. As shown in Figure 3 and Table 4, mortgage rate-adjustments had only a small impact on the HPI between March 2020 and December 2021; and for the year 2021, the adjustments would lead to an 3% increase in price reading, suggesting that the homel prices in 2021 would be even higher than had the mortgage rate impact been neutralized.

Figure 3. Mortgage-Rate-Adjusted Case-Shiller HPI During the Pandemic

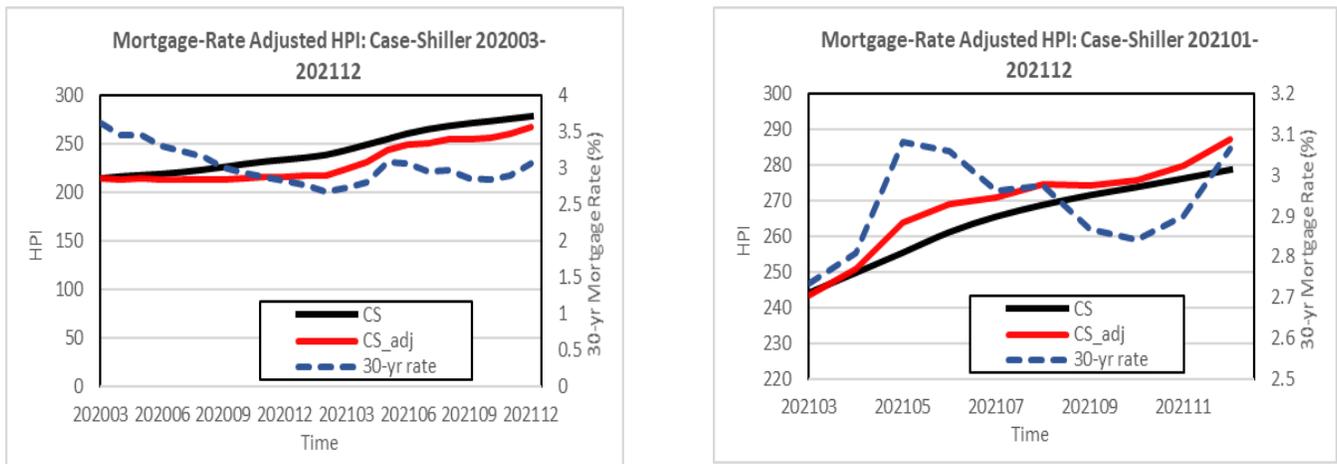

**Table 4. The Impact of Mortgage Rates on Case-Shiller HPI During the Pandemic**

| Timeline | Nominal Growth (A) | Adjusted Growth (B) | Impact from Mortgage Rates (C) |
|---|---|---|---|
| 202003-202112 | 29.5% | 24.4% | 3.9% |
| 202101-202112 | 17.8% | 21.4% | -3.0% |

Note: C = (A-B)/(1+A). The down payment rate is assumed as 20%.